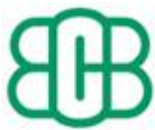

TRAJECTOIRE D'UN SATELLITE ARTIFICIEL

EN CHUTE AUTOUR

DE LA TERRE

par

Younes Ben-Ghabrit, Ali Lahlou Mimi

et Modar Hilmi Zawati

Travail présenté

à

Monsieur Stéphane Beauregard

dans le cadre du cours de

Calcul et intégration

201-BAC-05

Collège de Bois-de-Boulogne

Lundi 13 décembre 2010

# Résumé

Notre projet consiste à déterminer, mathématiquement, la trajectoire que prendrait un satellite artificiel en prenant en compte la résistance de l'air. Le but du travail est aussi de s'assurer que le satellite s'écrasera bel et bien sur la surface de la Terre. Pour commencer, nous avons étudié les forces mises en jeu entre un satellite quelconque et la Terre. Suivant cela, nous avons développé la deuxième loi de Newton en prenant en compte le frottement de l'air ainsi que la vitesse du satellite afin de trouver les équations reliant la trajectoire ainsi que le temps, la vitesse et la densité de l'air à une hauteur donnée. Aussi, il a fallu trouver une relation mathématique, grâce à des données expérimentales, de la densité en fonction de la hauteur du satellite afin d'inclure cette dernière dans l'équation du mouvement. Finalement, nous avons calculé le temps que prendrait le satellite afin de s'écraser sur Terre si nous diminuons sa vitesse initiale de 0 % à 12 % (situation réaliste). De façon à vérifier notre modèle, nous verrons ce qui arrivera si nous donnons une vitesse nulle au satellite.

Mots clés : orbite, densité, altitude, température, pression, vitesse, période.

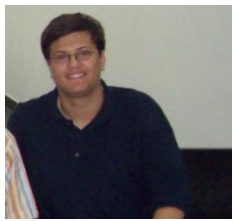
Younes Ben-Ghabrit
Ville Saint-Laurent
younesb91@gmail.com
Mathématiques pures et appliquées,
Université de Montréal

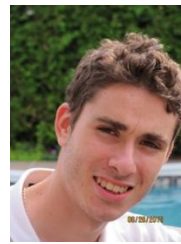
Ali Lahlou Mimi
Laval
bekham76@hotmail.com
Génie civil,
Université de Concordia

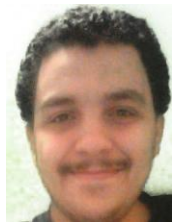
Modar Hilmi Zawati
Ville Saint-Laurent
modar90@gmail.com
Études cinématographiques,
Université de Concordia

# Abstract and keywords

The project consists to determine, mathematically, the trajectory that will take an artificial satellite to fight against the air resistance. During our work, we had to consider that our satellite will crash to the surface of our planet. We started our study by understanding the system of forces that are acting between our satellite and the earth. In this work, we had to study the second law of Newton by taking knowledge of the air friction, the speed of the satellite which helped us to find the equation that relates the trajectory of the satellite itself, its speed and the density of the air depending on the altitude. Finally, we had to find a mathematic relation that links the density with the altitude and then we had to put it into our movement equation. In order to verify our model, we'll see what happens if we give a zero velocity to the satellite.

Keywords: orbit, density, altitude, temperature, pressure, velocity, period.

# Table des matières

# Liste des symboles et des abréviations

| Symbole | Description | Unité | Symbole | Description | Unité |
|---|---|---|---|---|---|
| $\vec{F_f}$ | Vecteur de la force de friction | N | $h_0$ | Hauteur initiale du satellite | m |
| $\vec{F_g}$ | Vecteur de la force gravitationnelle | N | $v_0$ | Vitesse initiale du satellite | m/s |
| $C_x$ | Coefficient de frottement aérodynamique | - | $\nabla T$ | Gradient thermique | °C/km |
| $\rho(h)$ | Densité de l'air en fonction de la hauteur | g/cm³ | $T_0$ | Température initiale | K |
| $A$ | La surface projetée perpendiculairement au frottement de l'air. | m² | $\vec{r}$ | Vecteur entre la surface terrestre et le satellite | m |
| $G$ | Constante de gravitation universelle | m³ kg⁻¹ s⁻² | $g$ | Accélération gravitationnelle | m/s² |
| $M$ | Masse de la terre | kg | $\vec{v}$ | Vecteur vitesse du satellite | m/s |
| $R$ | Distance entre le centre de la terre et le satellite | m | $\vec{a}$ | Vecteur accélération satellite | m/s² |
| $x$ | Position du satellite en x | m | $y$ | Position du satellite en y | m |
| $r_T$ | Rayon de la terre | m | $x_0$ | Position initiale du satellite en x | m |
| $r$ | Distance entre la surface terrestre et le satellite | m | $y_0$ | Position initiale du satellite en y | m |
| $t$ | La durée de vol du satellite | s | $t_0$ | Temps initial | s |
| $T$ | Période du satellite | s | $M_{air}$ | Masse molaire de l'air | g/mol |

# Introduction

Pour commencer, étant donné l'importance de choisir un projet en bonne et due forme, nous avons opté pour un sujet touchant le domaine des satellites artificiels et de l'astrophysique. En effet, notre projet consiste à trouver la trajectoire de la chute d'un satellite artificiel gravitant initialement sur une orbite à une vitesse constante. Ainsi, pour débuter le processus de chute libre du satellite, il faut d'abord considérer le changement brusque d'orbite due à une modification de vitesse. Bien sûr, étant donné la complexité du projet, nous avons simplifié quelque peu celui-ci afin de respecter l'échéancier et ainsi, arriver à répondre à la problématique de départ. Nous avons premièrement décidé de ne pas calculer la valeur de la force appliquée au satellite pour lui faire changer d'orbite, aussi, nous n'avons pas considéré la force du vent appliquée sur le satellite dû à la très grande variation de celle-ci dans un temps déterminé.

Notre objectif dans ce projet est de parfaire nos connaissances acquises tout au long du programme de sciences de la nature et de les appliquer d'une façon efficace et pédagogique. Non seulement c'est une préparation pour les études supérieures, mais aussi, cela nous aide à développer une méthode de travail soignée et appliquée. De ce fait, respecter l'échéancier est une de nos priorités, car savoir gérer le temps nous servira tout au long de notre cheminement scolaire et professionnel. Bref, relever le défi de bien résoudre notre problématique initiale est notre principal objectif.

Pour y arriver, nous avons séparé la tâche dès le début en ciblant un aspect dans lequel chacun d'entre nous excelle. Par exemple, Modar a dû prendre l'initiative de calculer le développement de la deuxième loi de Newton pour arriver aux équations de cinématiques. Ali, quant à lui, s'est proposé de trouver une équation pour la variation de la densité de l'air en fonction de l'altitude. Finalement, Younes a calculé, avec la méthode d'Euler, l'approximation de la trajectoire du satellite et a effectué les simulations appropriées. Ainsi, nous y avons tous mis notre grain de sel.

Tout au long de ce projet, nous avons inclus toutes les parties expliquant notre cheminement étape par étape. En effet, nous avons divisé ce rapport en plusieurs aspects.

Premièrement, nous présenterons l'état de la recherche, soit un petit historique sur une problématique semblable à la nôtre. Deuxièmement, nous aborderons l'état de la recherche ainsi que les notions préalables qui nous ont servi à bien traiter nos calculs. Troisièmement, le développement se sépare en plusieurs sous-aspects. En effet, nous décomposerons et nous schématiserons le phénomène avec des cas plus simples. Ensuite, nous modéliserons notre problématique avec des équations plus rigoureuses que dans les cas simples. Après vient l'étape de simulation, c'est-à-dire l'étape cruciale dans notre cas, car c'est grâce à celle-ci que nous saurons si nos calculs sont fondés et adéquats, ou , par conséquent, erronés. Finalement, l'aspect de la problématique se soldera par un rapport de synthèse. Ainsi, nous terminerons l'analyse de ce rapport par une discussion qui viendra résumer tous les résultats obtenus ainsi qu'expliquer le fondement de ces derniers.

# L'état de la recherche (revue de la littérature)

À la lumière de nos recherches, nous avons constaté qu'au fil du temps, il y a eu beaucoup de situations contemporaines touchant de peu notre sujet.  Nous allons en mentionner quelques-unes.   Le vaisseau Soyouz-1, qui a été lancé le 23 avril 1967 et qui contenait le cosmonaute Vladimir Komarov avec son équipage, vit  un bon vol spatial, cependant vers la fin de la mission, le vaisseau a eu de la difficulté lors de sa rentrée atmosphérique et ses voyageurs ont perdu la vie vers la fin de cette mission.  En 1998, une capsule européenne de type Apollo fait une excellente rentrée atmosphérique, cette expérience avait pour but de faire uniquement une démonstration de cette rentrée. En 2003, sept astronautes, de la navette Columbia, sont décédés lors d'une rentrée atmosphérique. Finalement, les autorités américaines ont vécu un dilemme lorsqu' ils ont voulu faire descendre un satellite d'espionnage américain, car ils ont eu peur des dangers sur une population de la rentrée atmosphérique de ce dernier.

# Problématique

Actuellement, les satellites artificiels sont d'une importance capitale (aussi bien dans le domaine des télécommunications que dans le domaine de l'astrophysique). Leur utilisation croît de jour en jour. Les terminaux GPS, par exemple, démontrent la croissance de l'utilisation des satellites artificiels. Après un certain nombre d'années, ces satellites deviennent soit obsolètes ou encore, hors-service. Il est ainsi nécessaire de faire disparaitre ces derniers afin de libérer l'orbite qu'ils occupent. Le but de notre projet est de trouver la trajectoire d'un satellite artificiel qui lui permettrait de s'écraser sur terre si nous diminuons sa vitesse de 0% à 12%. Afin de vérifier si notre modèle est juste, nous donnerons une vitesse initiale nulle au satellite. Le calcul du temps total de la trajectoire du satellite est aussi un des buts du projet. Pour y arriver, plusieurs idées sont à formuler.

Tout d'abord, nous avions à considérer que, lors du déplacement du satellite vers la terre, la résistance de l'air est très présente et dépend de la hauteur de celui-ci. La hauteur du satellite peut donc être représentée par la densité de l'air. Ainsi, nous nous sommes demandés : «quelle est la relation entre la hauteur et la densité de l'air ?». Pour réussir à trouver cette relation, nous avons utilisé des données expérimentales pour modéliser cette situation.

Ainsi, après avoir obtenu une relation mathématique entre la hauteur et la densité de l'air, la prochaine étape serait donc de trouver la trajectoire du satellite en prenant en compte la résistance de l'air. Ainsi, notre second questionnement est le suivant: « Quelle est la trajectoire du satellite si nous considérons la friction de l'air ? ». Pour cela, nous avons étudié le système de forces mises en jeu entre le satellite et la terre.

Finalement, après avoir trouvé la trajectoire prise par le satellite, nous voulions savoir si une diminution de vitesse pouvait affecter sa trajectoire et donc, diminuer «la durée de vol» de celui-ci (le temps total autour de la terre). Ainsi, une troisième interrogation nous est ensuite venue à l'esprit : «Si nous diminuons la vitesse du satellite de 0% à 12 % de sa vitesse initiale, en combien de temps ce dernier s'écrasera-t-il sur la surface terrestre ?» Pour réussir à répondre à cette dernière interrogation, nous tracerons la trajectoire du satellite à toutes les diminutions de 3% de la vitesse initiale de celui-ci afin de calculer le temps pour chacun des intervalles.

# Évaluation de la démarche

Nous avons eu beaucoup d'obstacles lors de notre recherche.  Au début de la recherche, nous avons eu de la difficulté à visualiser la chute du satellite vers notre planète.  En effet, nous n'étions pas capables de schématiser la forme de la trajectoire qu'empruntera le satellite lors de sa chute vers la surface de la Terre.  Donc, nous sommes allés consulter un enseignant de physique, Monsieur René Lafrance, qui nous a illustré la situation d'une chute libre d'un satellite. Par ailleurs, nous avons eu à trouver la direction des vecteurs forces qui agissaient sur notre satellite et nous avons finalement choisi d'utiliser le système de coordonnées cartésiennes : x et y (vecteurs i et j). Aussi, comme nous avons eu de la difficulté à trouver des équations qui reliaient certaines de nos données, nous avons pris des données expérimentales.

Nous avons découvert, pendant ce projet, l'importance de la réflexion. En d'autres mots, essayer de comprendre en profondeur le sujet choisi, d'établir des liens entre ce dernier et les connaissances scientifiques déjà acquises.  Donc, nous avons aussi découvert l'utilité de ces apprentissages pour ce projet et pour ceux à venir dans nos vies respectives.  Nous avons aussi découvert que même une simple situation qui peut se décrire qualitativement en une phrase, peut s'exprimer difficilement quantitativement en plusieurs équations.

Dans ce projet, nous avons compris à quoi ressemblera la trajectoire du satellite. Cependant, nous avons eu de la difficulté à l'exprimer ou à la résoudre mathématiquement.  Nous nous sommes mis au travail pour trouver une façon de résoudre cette étape ultime du projet.  Tout d'abord, il est utile de bien diviser notre situation en plusieurs parties.  Comme le phénomène se révèle très complexe, nous commençons par le décortiquer par des cas plus simples. Ainsi, en comprenant des situations plus simples, nous incluons les facteurs qui risquent d'affecter celles-ci.

Ce projet nécessite beaucoup de réflexion. Afin d'être sur la bonne voie, il est utile de mettre en œuvre un protocole qui explique les étapes à suivre pour notre projet.  En effet, la première étape, qui est la problématique en question, nous permet d'identifier le problème et de se poser des questions auxquelles nous allons répondre pendant notre recherche. Nous incluons aussi une hypothèse à notre planification (cette deuxième étape nous a donné la chance de trouver des prédictions de ce qui arrivera au début, pendant et vers la fin de notre projet). La troisième étape est d'identifier les sources disponibles pouvant nous aider au fur et à mesure de l'élaboration de la solution. Cette étape est importante, car elle nous facilite une bonne partie de la recherche. Ensuite, la quatrième étape est une étape incontournable dû au fait qu'elle nous aide à développer nos idées. Nous écrivons ces dernières dans un cahier de bord. Cette étape est cruciale, car nos idées doivent être bien claires. Nous nous en servirons d'ailleurs comme une bonne base pour la suite du projet. La cinquième étape est nécessaire, car elle nous permet de valider nos démarches et lorsqu'il y a des erreurs, nous avons la possibilité de les corriger. Finalement, la dernière étape quant à elle est très utile, car c'est la résolution de notre problématique, bref la solution du projet en soi.

Il est indispensable d'avoir une bonne communication entre les membres de l'équipe. Aussi, ne pas s'immobiliser sur un problème est essentiel, parce que cela nous évite de gaspiller un temps précieux et d'essayer de trouver une solution à ce dernier. Par ailleurs, il faut aussi être capable d'exprimer nos idées avec des relations mathématiques qui seront utiles plus tard pour résoudre notre problématique.

Pour s'assurer que nos progrès sont réels, nous revenons souvent à nos démarches et nous essayons de faire le lien entre le travail déjà fait et le travail à accomplir. Nous communiquons entre nous pour valider nos équations, nos idées et nos remarques. Nous consultons souvent internet et les manuels de physique pour en tirer des connaissances approfondies qui nous serviront par la suite. Finalement, Monsieur Beauregard, professeur de mathématiques ainsi que Monsieur Lafrance, le professeur de physique ont pu nous donner quelques pistes à suivre pour mieux nous guider tout au long du projet.

# Notions théoriques préalables

**Transformation des coordonnées cartésiennes en coordonnées polaires :**

Comme notions préalables, les coordonnées polaires sont d'une importance capitale car nous les utiliserons lors de la détermination de la trajectoire du satellite.

Considérons le point (x,y) suivant :

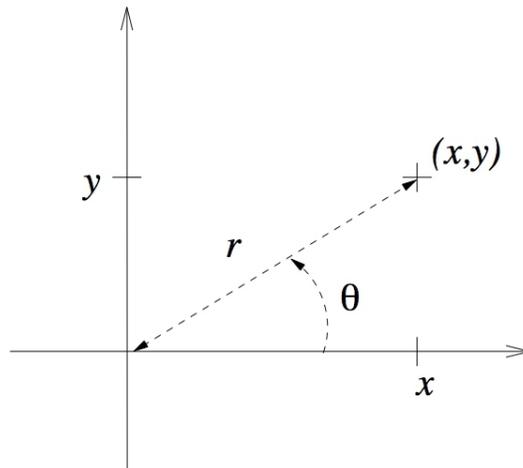

Wikipedia ©

En coordonnées cartésiennes, nous pouvons exprimer l'angle θ comme étant :

$$\text{Arctan} \frac{y}{x} = \theta$$

Si nous voulons calculer la valeur du rayon, nous pouvons utiliser le théorème de Pythagore :

$$r = \sqrt{x^2 + y^2}$$

Nous pouvons ainsi dire que $r^2 = x^2 + y^2$

Avec ce qui a été formulé précédemment, nous pouvons dire que

$$x = r\cos(\theta)$$
$$y = r\sin(\theta)$$

## Résolution d'équations différentielles à variables séparables

Aussi, comme connaissance préalable, nous ne pouvons faire notre projet si la résolution d'équations différentielles à variables séparables n'est pas chose acquise. Nous utiliserons donc cette méthode afin de trouver deux équation de cinématique (découlant de la deuxième loi de Newton) et aussi, de trouver une relation entre la pression et la hauteur du satellite.

Supposons que l'équation différentielle suivante peut être écrite pour tous les x sous la forme :

$$\frac{d}{dx}f(x) = g(x)h(f(x)), \qquad (1)$$

Une écriture plus simple peut être exprimée comme voici :

$$\frac{dy}{dx} = g(x)h(y) \qquad (1)$$

Si h(y) ≠ 0, nous pouvons réécrire les termes de l'équation pour obtenir :

$$\frac{dy}{h(y)} = g(x)dx,$$

En séparant donc x et y.

## Méthode d'Euler

Finalement, nous avons utilisé la méthode d'Euler pour résoudre deux équations différentielles pour lesquelles la résolution n'était pas possible avec les méthodes directes connues.

Soit l'équation différentielle du 1$^{er}$ ordre :

$$y' = f(x, y)$$
$$y(x_0) = y_0$$

Si $x_1 = x_0 + h$ (où $h \in \mathbb{R}$) est suffisamment près de $x_0$ et que y(x) peut être obtenue par son développement en série de Taylor autour de x=$x_0$

$$y(x_1) = y_0 + (x_1 - x_0)y'(x_0) + \frac{(x_1 - x_0)^2}{2!}y''(x_0) + \cdots + \frac{(x_1 - x_0)^n}{n!}y^{(n)}(x_0)$$

$$= y_0 + hy'(x_0) + \frac{h^2 y''(x_0)}{2!} + \cdots + \frac{h^n y^{(n)}(x_0)}{n!} + \cdots$$

Puisque c'est une série convergente, nous savons que les termes de la série $\frac{h^n y^{(n)}(x_0)}{n!} \rightarrow$ 0, lorsque $n \rightarrow \infty$.

En gardant les n premiers termes et en négligeant les autres, on peut démontrer que l'erreur commise ne dépasse pas $\frac{h^{n+1} y^{(n+1)}(\xi)}{(n+1)!}$, où $\xi$ est un point de l'intervalle $[x_0, x_1]$

Ainsi, si h est suffisamment petit, on peu minimiser l'erreur à volonté. En conservant les deux premiers termes, nous obtenons :

$$y(x_1) = y_0 + hy'(x_0)$$

# Développement

Décomposition du phénomène

Tout d'abord, afin de mieux comprendre comment trouver la trajectoire empruntée par le satellite, nous avons commencé par simplifier la situation en prenant des cas simples de phénomènes physiques comme par exemple la chute libre d'un objet ayant seule la gravitation comme facteur de force. Ainsi, le but était de calculer la position et la vitesse exacte à un moment précis. Pour y arriver, nous avons fait appel à nos connaissances de physique mécanique en utilisant la 2$^e$ loi de Newton qui est la suivante : $\vec{F} = m\vec{a}$. Ainsi, nous avons développé quelque peu cette formule en utilisant des intégrales pour arriver à ce que nous appelons les équations de cinématiques. Voici la preuve :

$$\sum \vec{F} = m\vec{a}$$
$$m\vec{g} = m\vec{a}$$

$$\vec{g} = \vec{a}$$

$$\frac{d\vec{v}}{dt} = \vec{g}$$

$$d\vec{v} = \vec{g}\, dt$$

$$\int d\vec{v} = \int \vec{g}\, dt$$

$$\int d\vec{v} = \vec{g} \int dt$$

$$\vec{v} + C_1 = \vec{g}t + C_2$$

$$\vec{v} = \vec{g}t + C \quad \text{où } C_2 - C_1 = C$$

Pour trouver la valeur de la constante, nous poserons la situation initiale où, à 0 s, la vitesse de l'objet en question est une vitesse initiale :

$$\vec{v_0} = 0\vec{g} + C$$

$$\vec{v_0} = C$$

$$\vec{v} = \vec{g}t + \vec{v_0}$$

$$\frac{d\vec{x}}{dt} = \vec{g}t + \vec{v_0}$$

$$d\vec{x} = (\vec{g}t + \vec{v_0})dt$$

$$\int d\vec{x} = \int (\vec{g}t + \vec{v_0})dt$$

$$\vec{x} + C_3 = \vec{g}t^2 + \vec{v_0}t + C_4$$

$$\vec{x} = \vec{g}t^2 + \vec{v_0}t + C \quad \text{où } C_4 - C_3 = C$$

☒

Pour trouver la valeur de la constante, encore une fois, nous poserons la situation initiale où à un temps nul, la position devient donc la position initiale.

D'où nous obtenons les formules suivantes :

$$v = at + v_0 \quad \text{et} \quad x_0 = \frac{a}{2}t^2 + v_0 t + x_0$$

Grâce à ces formules, nous étions capable de calculer la position exacte ainsi que la vitesse d'un objet en chute libre ou encore un objet se déplaçant sur deux dimensions.

Ensuite, pour relater notre projet à cette situation de simple chute libre, nous avons introduit la notion de force de frottement.

En effet, cette force s'explique par le fait que l'objet est ralenti par une force qui va à l'encontre du mouvement. Donc nous obtenons la formule suivante pour le calcul des forces :

$$\vec{F}_f + \vec{F}_g = m\vec{a} \qquad (1.0)$$

Voilà un schéma simple des forces impliquées :

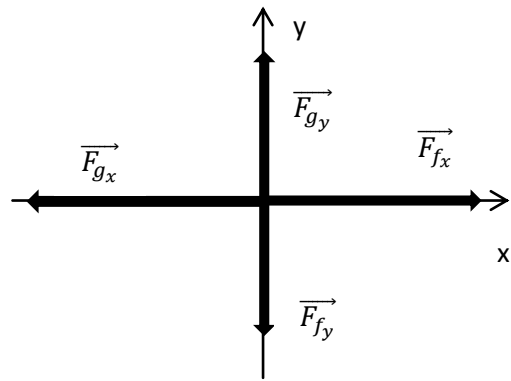

Figure 1

Celui-là représente le schéma des forces en présence les masses:

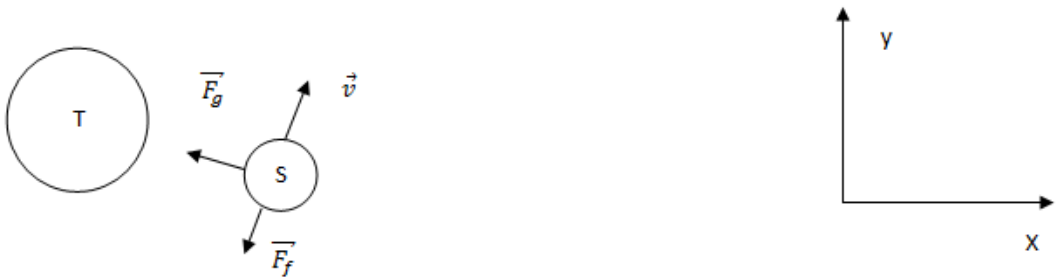

Figure 2

Cependant, pour adapter cette force à notre situation, il a fallu trouver l'équation d'une force de frottement variable, puisque la densité de l'air est variable selon l'altitude. Ainsi, en effectuant quelques recherches, nous avons pu établir une relation entre la force de frottement et la force de traînée. Cette relation se nomme la trainée de frottement. La force de traînée s'exprime comme suit :

$$\vec{F}_f = \frac{1}{2} C_x \rho A \vec{v} |\vec{v}|,$$

où $C_x$, $\rho$, $A$ et $v$ sont, respectivement le coefficient de traînée, la densité de l'air, la surface projetée perpendiculairement à la force de frottement (nous considérons que le satellite est une sphère uniforme de rayon 50 m) et la vitesse de l'objet.

D'autre part, comme notre situation s'agit d'un cas où la gravité n'est pas constante, en d'autres mots, elle varie selon l'altitude, nous ne pouvons continuer à nous servir des équations de cinématiques, car notre satellite se situe initialement à 600 Km de la Terre.

Puisqu'il s'agit d'un cas où la gravitation terrestre est très importante, il a fallu faire appel à l'équation de la loi gravitation universelle : $\vec{F}_g = \frac{GMm}{\vec{R}^3} \vec{R}$, où $G$, $M$, $m$, $R$ sont, respectivement, la constante gravitationnelle, la masse de la Terre, la masse de l'objet gravitant autour de la Terre et le rayon de la Terre.

Donc, nous obtenons la formule suivante :

$$\frac{1}{2} C_x \rho A \vec{v} |\vec{v}| + \frac{GMm}{\vec{R}^3} \vec{R} = m\vec{a} \qquad (1.1)$$

Par la suite, nous devons nous pencher sur l'énergie mécanique du satellite, car nous en avons besoin pour les calculs de variation de la vitesse suite aux changements d'orbites. De ce fait, l'énergie mécanique totale est l'addition de l'énergie potentielle de gravitation et de l'énergie cinétique du corps de masse $m$ gravitant autour de la Terre. La formule s'écrit ainsi :

$$E = \frac{1}{2} mv^2 - \frac{GMm}{R}, \qquad (1.2)$$

où R est la distance entre les deux corps en mouvement.

Pour ce qui en est de la trajectoire du satellite, après quelques recherches, nous sommes arrivés à la conclusion que si nous voulons faire atterrir le satellite sur terre, nous devons faire diminuer sa vitesse. Ainsi, cette variation de vitesse sera choisie par tâtonnements sur une feuille excel pour ainsi voir quand le satellite s'écrasera (on choisira cette vitesse quand il sera entre une hauteur positive et une hauteur négative). Nous aurons ainsi, le temps nécessaire pour l'écrasement du satellite, sa vitesse, sa hauteur et sa position. Voici un schéma décrivant cette situation en passant d'une orbite circulaire vers une orbite elliptique ayant comme extrémité d'un des grands rayons : la Terre.

Voici un schéma représentant la situation[1] :

**Image d'une simulation animée d'une trajectoire
d'un satellite artificiel se dirigeant vers la terre**

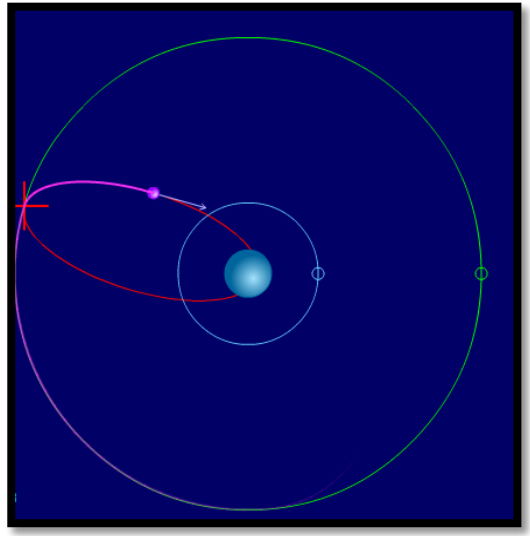

---

[1] *http://www.sciences.univ-nantes.fr/physique/perso/gtulloue/Meca/Planetes/transfert.html*

## Construction de la solution

Tout d'abord, nous allons étudier la variation de densité afin de pouvoir avoir une relation en fonction de la hauteur du satellite artificiel.

Pour mieux comprendre le phénomène de la variation de la densité, nous prenons la loi des gaz parfaits et nous isolons la densité, ce qui nous donne :

$$\rho(h) = \frac{M_{air}p(h)}{RT(h)} \quad (1.3)$$

où $M, p(h), R$ et $T(h)$ sont, respectivement, la masse molaire de l'air, la pression en fonction de la hauteur, la constante des gaz parfaits et la température en fonction de la hauteur.

Pour trouver la relation entre la pression en fonction de la hauteur, nous développons quelques calculs à l'aide de quelques notions théoriques :

$$\frac{dp}{dh} = \frac{-pM_{air}g}{RT}$$

Nous commençons par énoncer l'équation différentielle à variables séparables

$$dp = \frac{-pM_{air}g}{RT}dh$$

Nous posons par la suite les intégrales de chaque côté de l'égalité puis on intègre

$$\int \frac{dp}{p} = -\int \frac{M_{air}g}{RT}dh$$

$$\ln|p| = \frac{M_{air}g}{RT} \int dh$$

$$\ln|p| = \frac{M_{air}gh}{R(T_h - \nabla T)},$$

où $T_h$ est la température à une certaine hauteur et $\nabla T$ est le gradient thermique

$$\ln |p| = \frac{M_{air}gh}{(RT_h - R\nabla T)}$$

On isole ensuite p

$$\ln |p| (RT_h - R\nabla T) = M_{air}gh$$

$$RT_h \ln |p| - R\nabla T \ln |p| = M_{air}gh$$

$$\ln|p| (T_h - R\nabla T) = M_{air}gh$$

$$\ln|p| = \frac{M_{air}gh}{(T_h - R\nabla T)}$$

où $T_h$ est la température à une certaine hauteur et $\nabla T$ est le gradient thermique

$$\ln |p| = \frac{M_{air}gh}{(RT_h - R\nabla T)}$$

On isole p et nous obtenons finalement,

$$p = e^{\frac{M_{air}gh}{(RT_h - R\nabla T)}}$$

⊠

Bref, pour ce qui est de l'équation finale de la densité en fonction de la hauteur, cela nous donne :

$$\rho(h) = -\frac{e^{M_{air}gh/(RT_h - R\nabla T)} M_{air}}{(RT_h - R\nabla T)}$$

Grâce à cette formule, nous pouvons calculer la densité de l'air, par conséquent la force de frottement, à n'importe quelle hauteur.

Si nous représentons cette variation graphiquement, nous obtenons ceci : [2]

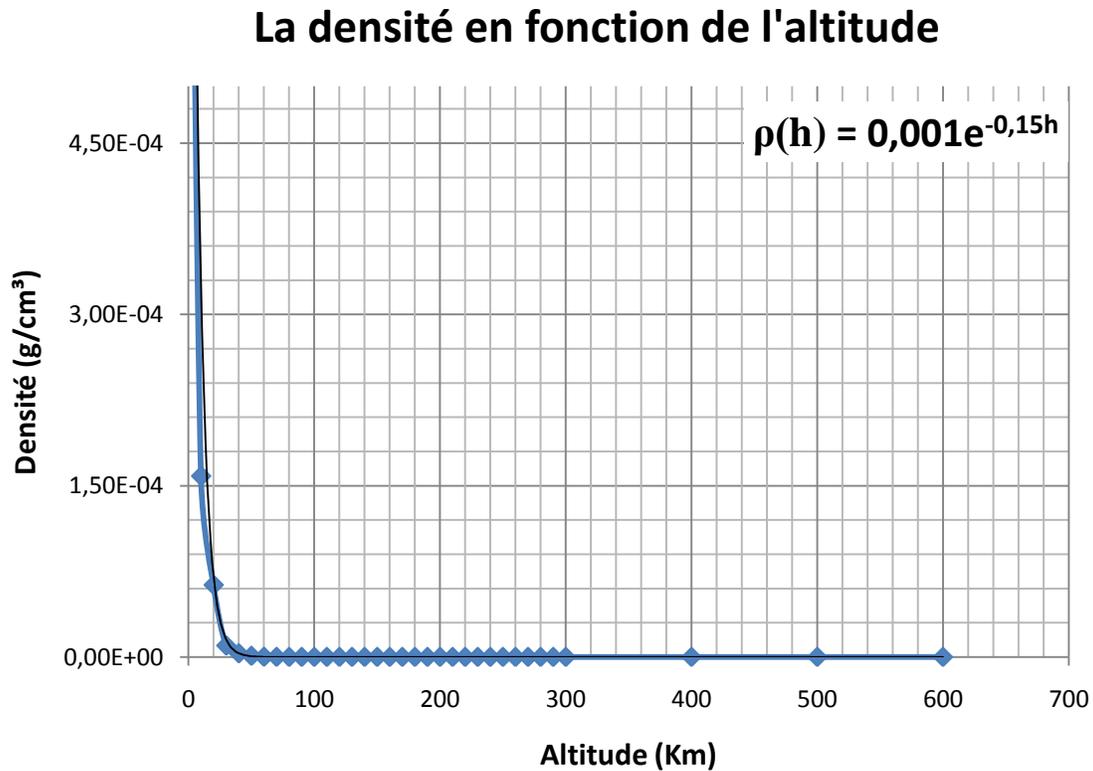

D'autres part, nous pouvons trouver la vitesse du satellite sur l'orbite circulaire (et par conséquent, sur l'extrémité horizontale de l'ellipse), pour cela, nous égalisons la force centripète à la loi de la gravitation universelle de Newton :

$$F_c = \frac{mv^2}{r} = \frac{GMm}{r^2},$$

où *r* est la distance entre le centre de la terre et le satellite.

$$v^2 = \frac{GM}{r}$$

---

[2] http://fr.wikipedia.org/wiki/Fichier:Atmosph%C3%A4re_Dichte_600km.png

$$v = \sqrt{\frac{GM}{r}} \tag{2.1}$$

La formule de la vitesse, par conséquent, ne prend pas en compte la masse du satellite.

Suite à cela, il faudra calculer la valeur de la vitesse initiale, qui est en fait la vitesse au début du cycle de l'orbite elliptique. Pour la calculer, il faut utiliser l'énergie mécanique du satellite. Voici les étapes de développement pour trouver la valeur de $v_0$ :

Nous savons que $E = \frac{GMm}{2R}$, où R est la distance entre les le centre de la Terre et le satellite. Ainsi, nous pouvons affirmer que $E = \frac{GMm}{2(R_T+R_{h2})}$, où $R_T$ est le rayon de la Terre et $R_{h2}$ est le rayon de l'orbite circulaire.

Donc $v_0 = \sqrt{\frac{2}{m}E}$ , qui est donc équivalent à $v_0 = \sqrt{\frac{GM}{r}}$

$$\Rightarrow v_0 = \sqrt{\left(\frac{(3,986\times 10^{14})}{(6371000+600000)}\right)}$$

En remplaçant, nous trouvons que $v_0 = 7561,72 \frac{m}{s}$

La formule de la vitesse orbitale, par conséquent, ne prend pas en compte la masse du satellite.

L'étude des forces liées au mouvement du satellite, selon la deuxième loi de Newton, en prenant compte de la friction de l'air est l'étape qui suit la simulation sans conditions (telle que la friction entre le satellite et l'air).

Étant donné la figure 2 :

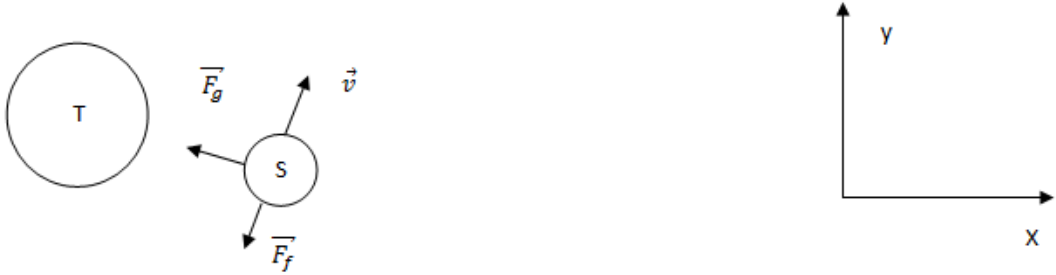

Figure 2

Posons :

$$\vec{r} = (x, y)$$

$$\vec{v} = (v_x, v_y)$$

$$\vec{a} = (a_x, a_y)$$

$$\vec{F_g} = (-m\vec{g}\frac{x}{\sqrt{x^2+y^2}}, -m\vec{g}\frac{y}{\sqrt{x^2+y^2}})$$

$$\vec{F_f} = (-\frac{1}{2}\rho(h)AC_x v_x|v_x|, -\frac{1}{2}\rho(h)AC_x v_y|v_y|)$$

Nous avons considérer le carré des composantes des vitesses en décomposant chacune des composantes en : $v_x{}^2 = v_x|v_x|$, $v_y{}^2 = v_y|v_y|$ car nous ne voulons pas la distance entre le satellite et la terre soit supérieure à celle donnée initialement (si nous avions $v_x{}^2 = v_x v_x$, nous avons la possibilité que $v_x$ soit négatif et donc, que la distance entre le satellite et la terre soit supérieure à la valeur initiale.

Nous pouvons ainsi obtenir la somme des forces selon les composantes x et y :

$$\sum \vec{F_x} = m\vec{g}\frac{x}{\sqrt{x^2+y^2}} + \frac{1}{2}\rho(h)AC_x\vec{v_x}|v_x| = m\vec{a_x}$$

$$\sum \vec{F_y} = m\vec{g}\frac{y}{\sqrt{x^2+y^2}} + \frac{1}{2}\rho(h)AC_x\vec{v_y}|v_y| = m\vec{a_y}$$

Nous avons modifié la formule de la force gravitationnelle pour chacune des composantes en multipliant cette dernière par : $\frac{x}{\sqrt{x^2+y^2}}$ et $\frac{y}{\sqrt{x^2+y^2}}$ (selon la composante x ou y) car nous voulons la force gravitationnelle selon une direction donnée (dans ce cas-ci, selon la direction x et y). En d'autres mots, nous avons multiplié les forces gravitationnelles respectives par $\frac{r\cos\theta}{r}$ et par $\frac{r\sin\theta}{r}$. Donc, nous avons multiplié par $\cos\theta$ (en x) et par $\sin\theta$ (en y) pour ainsi obtenir la composante dans la direction souhaitée.

Nous pouvons obtenu l'accélération en divisant la somme des forces de chacune des composantes par la masse du satellite :

$$\vec{a}_x = \vec{g}\frac{x}{\sqrt{x^2+y^2}} + \frac{1}{2m}\rho(h)AC_x v_x|v_x|$$

$$\vec{a}_y = \vec{g}\frac{y}{\sqrt{x^2+y^2}} + \frac{1}{2m}\rho(h)AC_x v_y|v_y|$$

Suite à cela, nous pouvons savons que $a = \frac{dv}{dt}$

$$\frac{dv_x}{dt} = -\vec{g}\frac{x}{\sqrt{x^2+y^2}} - \frac{1}{2m}\rho(h)AC_x v_x|v_x| \tag{2.2}$$

$$\frac{dv_y}{dt} = -\vec{g}\frac{y}{\sqrt{x^2+y^2}} - \frac{1}{2m}\rho(h)AC_x v_y|v_y| \tag{2.3}$$

Ces équations différentielles ne peuvent pas se résoudre avec une méthode standard de résolution d'équations différentielles. Nous avons donc utilisé la méthode d'Euler pour avoir une bonne approximation de cette dite équation. Nous avons tout d'abord, pour bien maîtriser la situation, considérer que le frottement était nul pour toutes les hauteurs. Les équations différentielles deviennent donc :

$$\frac{dv_x}{dt} = -\vec{g}\frac{x}{\sqrt{x^2+y^2}} \tag{2.4}$$

$$\frac{dv_y}{dt} = -\vec{g}\frac{y}{\sqrt{x^2+y^2}} \tag{2.5}$$

Nous avons décidé, afin d'améliorer légèrement la précision de notre modélisation, de calculer l'accélération gravitationnelle à chacune des hauteurs :

$$g(r) = -\frac{GM}{r} \tag{2.6}$$

Nous obtenons ainsi la vitesse du satellite s'il suivait une trajectoire circulaire (mouvement circulaire uniforme). Ainsi, le modèle précédent est valide si nous considérons que la terre est uniformément circulaire (que son rayon ne varie pas) et qu'il n'y a pas de trous d'air. Nous utiliserons donc le modèle précédent en diminuant petit à petit la vitesse du satellite afin de pouvoir constater l'effet de ces modifications de vitesses sur la trajectoire de ce dernier. Ainsi, il sera questions de diminuer la vitesse de la trajectoire avec frottement du satellite. Finalement, nous allons vérifier si notre modèle de la trajectoire du satellite avec frottement est plausible en diminuant la vitesse initiale du satellite à 0 m/s.

## Étude et simulation

Conditions initiales :

| Symbole | Valeur |
|---------|--------|
| $v_0$ | 7570,36 m/s |
| $v_{0x}$ | 0 m/s |
| $v_{0y}$ | 7561,72 m/s |
| m | 11 000 kg |
| M | $5,97 \times 10^{24}$ |
| $h_0$ | 600 000 m |
| G | $6,67 \times 10^{-11}$ m³/s.kg |
| $r_T$ | 6 371 000 m |
| T | 5906,66 s |

**1)** Étude de la trajectoire du satellite en orbite avec et sans frottement

Tout d'abord, simulons la trajectoire du satellite en considérant qu'il n y ait pas de frottement avec l'air. Nous devons tout d'abord résoudre les équations différentielles (2.4) et (2.5) par la méthode d'Euler (car, comme dit précédemment, il n y a pas de méthode directe pour résoudre de genre d'équations différentielles). Nous poserons comme valeurs initiales (comme dit précédemment) :

$$v_{0y} = 7561{,}72 \frac{m}{s} \ , \ v_{0x} = 0 \frac{m}{s}$$

et

$$x_0 = r_T + r = 6\ 356\ 752\ m + 600\ 000\ m = 6\ 956\ 752\ m$$

Prenons un pas de $\Delta t = 1{,}8$ secondes.

Nous obtenons ainsi pour x :

$$v_{0x} = 0$$
$$v_{1x} = 0 + 1{,}8 \left( g \left( \frac{x_0}{\sqrt{x_0^2 + y_0^2}} \right) \right)$$
$$v_{2x} = 1{,}8 \left( g \left( \frac{x_0}{\sqrt{x_0^2 + y_0^2}} \right) \right) + 1{,}8 \left( g \left( \frac{x_1}{\sqrt{x_1^2 + y_1^2}} \right) \right)$$

Et ainsi de suite.

Pour y, le principe est le même :

$$v_{0y} = 7561,72$$
$$v_{1y} = 7561,72 + 1,8(g(\frac{y_0}{\sqrt{x_0^2 + y_0^2}}))$$
$$v_{2y} = 7561,72 + 1,8(g(\frac{y_0}{\sqrt{x_0^2 + y_0^2}})) + 1,8(g(\frac{y_1}{\sqrt{x_1^2 + y_1^2}}))$$

Et ainsi de suite. Graphiquement, pour cette orbite stable, nous obtenons la trajectoire suivante :

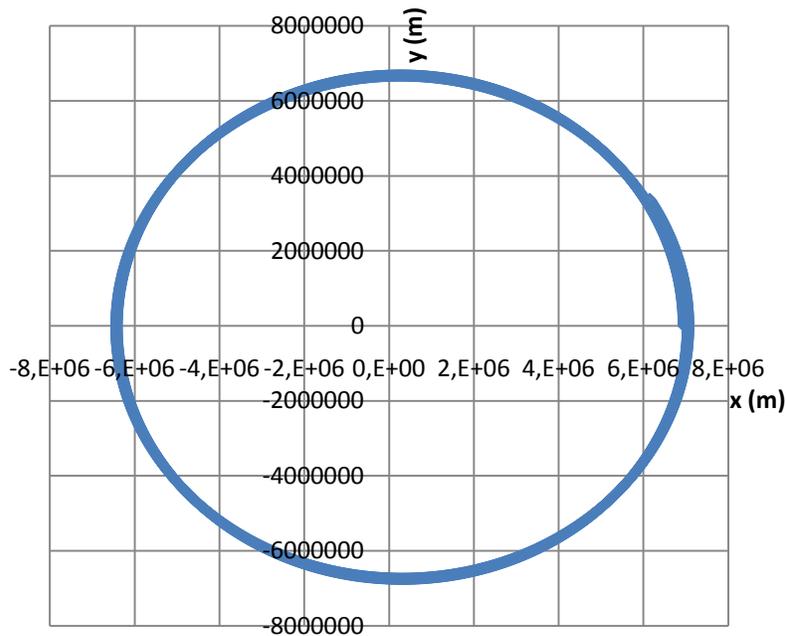

**Position en x et en y d'un satellite en orbite stable autour de la terre**

Maintenant que nous savons que notre équation est juste, nous pouvons modéliser la trajectoire avec frottement. Pour cela, nous résoudrons les équations (2.2) et (2.3) grâce à la méthode d'Euler. Ainsi, nous prendrons les mêmes valeurs initiales que pour le cas sans frottement soit :

$v_{0y} = 7561,72 \frac{m}{s}$ , $v_{0x} = 0 \frac{m}{s}$

et

$$x_0 = r_T + r = 6\,356\,752\,m + 600\,000\,m = 6\,956\,752\,m$$

Ainsi, la résolution se fait comme lors de la trajectoire sans friction en ajoutant, bien entendu, la friction :

$$v_{0x} = 0$$

$$v_{1x} = 0 + 1{,}8\left(-g\frac{x_0}{\sqrt{x_0^2 + y_0^2}} - \frac{1}{2m}\rho(h)AC_x v_{0x}|v_{0x}|\right)$$

$$v_{2x} = 1{,}8\left(-g\frac{x_{0x}}{\sqrt{x_{0x}^2 + y_{0x}^2}} - \frac{1}{2m}\rho(h)AC_x v_{0x}|v_{0x}|\right) + 1{,}8(-g\frac{x_{1x}}{\sqrt{x_{1x}^2 + y_{1x}^2}} - \frac{1}{2m}\rho(h)AC_x v_{1x}|v_{1x}|$$

Et ainsi de suite.

Pour la composante y, la manière de procéder est la même.

$$v_{0y} = 7561{,}72$$

$$v_{1y} = 7561{,}72 + 1{,}8\left(-g\frac{y_0}{\sqrt{x_0^2 + y_0^2}} - \frac{1}{2m}\rho(h)AC_x v_{0y}|v_{0y}|\right)$$

$$v_{2y} = 7561{,}72 + 1{,}8\left(-g\frac{y_{0x}}{\sqrt{x_{0x}^2 + y_{0x}^2}} - \frac{1}{2m}\rho(h)AC_x v_{0y}|v_{0y}|\right) + 1{,}8(-g\frac{y_{1x}}{\sqrt{x_{1x}^2 + y_{1x}^2}} - \frac{1}{2m}\rho(h)AC_x v_{1y}|v_{1y}|)$$

Et ainsi de suite

Graphiquement, nous obtenons ceci :

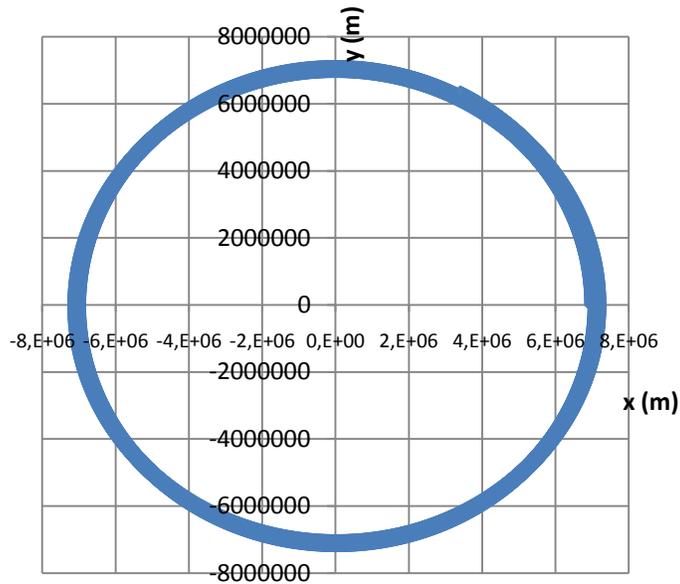

**Position en x et en y d'un satellite en orbite stable autour de la terre**

Nous avons maintenant le graphique de la trajectoire du satellite si nous considérons que le frottement du satellite avec l'air ainsi que l'accélération gravitationnelle varient selon la hauteur de celui-ci.

2) Diminution de la vitesse du satellite

Afin de faire atterrir le satellite, une diminution de vitesse de celui-ci est inévitable. Pour cela, nous allons procéder à une diminution graduelle de la vitesse du satellite de 3 % jusqu'à 12 %. En d'autres mots, nous passerons de $v_{orb} = 7561{,}72 \frac{m}{s}$ à $v_0 = 7334{,}87 \frac{m}{s}$ et nous allons tracer la trajectoire du satellite pour ainsi observer l'atterrissage de celui-ci.

Posons $v_0 = 7334{,}87 \frac{m}{s}$

La trajectoire du satellite devient :

**Trajectoire du satellite en chute autour de la terre**

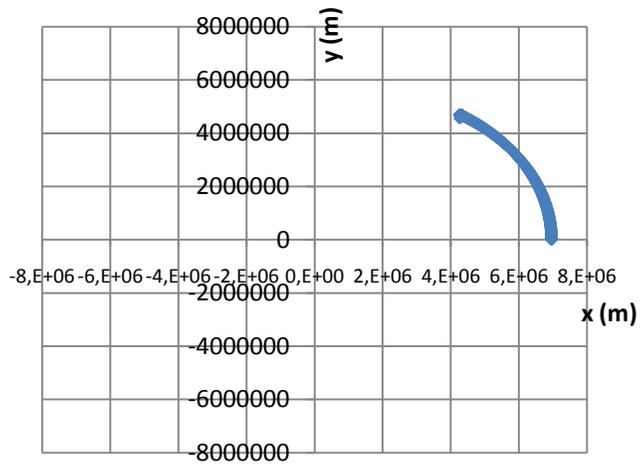

Nous pouvons donc voir que le satellite s'écrase sur la surface de la terre de manière graphique. D'une manière numérique, il est aussi possible de déduire cela en voyant la hauteur du satellite devenir soudainement négative. Voici cette variation mise en tableau :

| t (s)    | h (m)       |
|----------|-------------|
| 1800,05  | 2458,6624   |
| 1801,9   | 1996,77221  |
| 1803,75  | 1561,20827  |
| 1805,6   | 1147,10622  |
| 1807,45  | 750,716485  |
| 1809,3   | 369,099156  |
| 1811,15  | -0,0812771  |
| 1813     | -358,69903  |

Nous pouvons donc conclure, pour ce cas précis, que le satellite atteindra la surface terrestre en 1811,15 s environ.

Voyons maintenant la trajectoire du satellite pour $v_0 = 7108,02\ \frac{m}{s}$ (une vitesse de 3 % inférieure à la précédente).

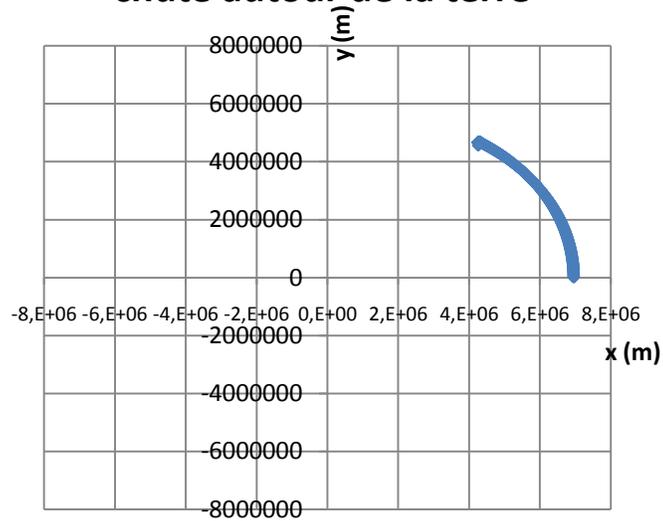

Trajectoire du satellite en chute autour de la terre

Nous trouvons ainsi que le satellite s'écrase sur terre en 1309,8 s par la même méthode utilisée précédemment. Nous remarquons que le satellite s'écrase sur terre plus rapidement que pour le changement de vitesse précédent.

Pour $v_0 = 6881,17\ \frac{m}{s}$, la trajectoire du satellite est la suivante :

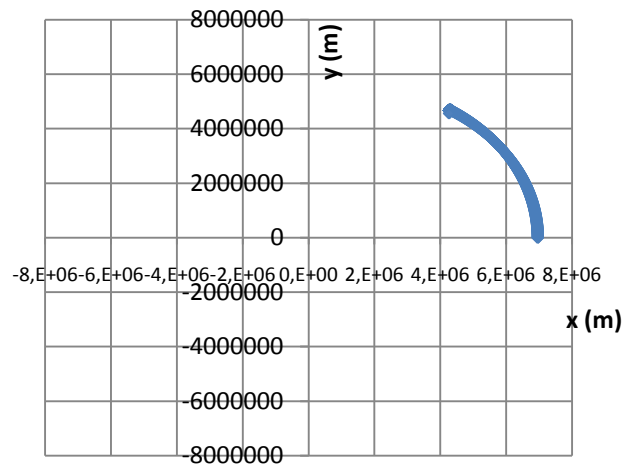

Trajectoire du satellite en chute autour de la terre

Nous trouvons ainsi que le satellite s'écrase sur terre en 1074,85 s. Pour la dernière valeur, nous prenons : $v_0 = 6654,31\ \frac{m}{s}$. La trajectoire est ainsi la suivante :

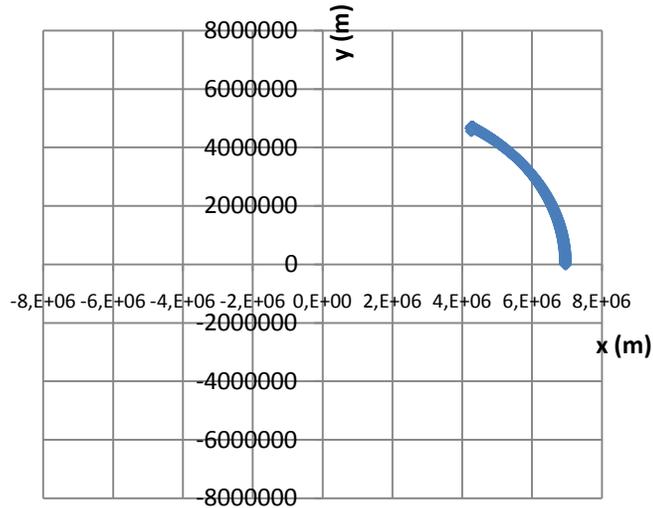

Il effectue ainsi sa collision avec la terre en 900,95 s.

Par rapport à la réalité, le modèle proposé ci-haut permettrait donc de libérer une orbite en une demi-heure au maximum (pour une diminution de vitesse orbitale de 3%.

Choisissons maintenant une vitesse initiale de 0 m/s ;

Voyons ce que cela donne graphiquement :

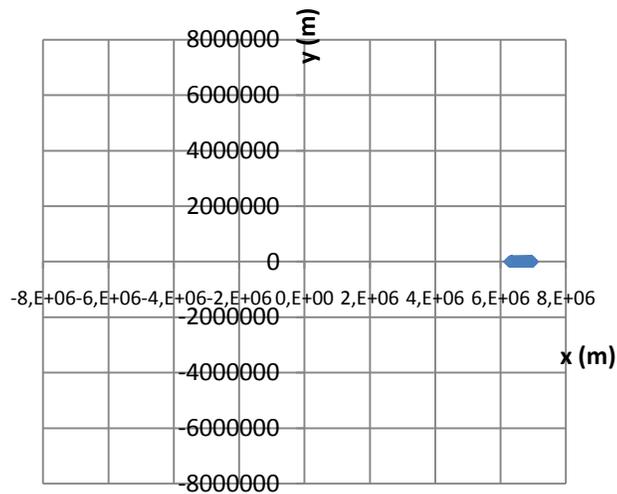

**Trajectoire du satellite en chute autour de la terre**

Nous remarquons que le satellite ne fait aucune parcelle de trajectoire elliptique. La trajectoire subie est en fait une chute libre de 600 000 m sans fluctuations en 416,25 s. La cause est que si un objet ayant une vitesse nulle est déployé d'une orbite stable, il ne subit plus de force centripète et donc, subit totalement l'attraction terrestre.

## Rapport de synthèse

Tout d'abord, nous avons trouvé (par simulation), le temps que prendrait un satellite artificiel pour atteindre la surface terrestre en modifiant sa vitesse orbitale. Dans un premier cas, nous avons diminué la vitesse initiale du satellite de 3% (7334,87 $\frac{m}{s}$) et nous avons obtenus une trajectoire totalisant un temps de 1811,15 s environ. Nous observons donc que le satellite s'écrase sur terre dans le premier cadran du graphique. En diminuant la vitesse orbitale de 6%, nous obtenons une vitesse de 7108,02 $\frac{m}{s}$ et un temps de trajectoire de 1309,8 s. Le satellite s'écrase aussi dans le premier cadran du graphique. Pour ce qui en est des diminutions de 9 % et 12 % de la vitesse orbitale initiale, nous obtenons une vitesse de 6881,17 $\frac{m}{s}$ et un temps d'écrasement de 1074,85 s pour la diminution de 9 % de la vitesse orbitale et une vitesse de 6654,31 $\frac{m}{s}$ en 900,95 s. Nous remarquons que ces deux dernières données (tout comme les autres trajectoires citées précédemment) se situent dans le premier cadran des graphiques de leur trajectoire respectives. Nous avons finalement simuler une trajectoire ayant une vitesse de $0\,\frac{m}{s}$. Nous nous sommes rendu compte que cela revenait à effectuer une chute libre de 600 000 m. Cette dite chute se produit donc en 416,25 s. Nous avons ainsi trouvé une raison pour laquelle le satellite s'écrasait si nous diminuons sa vitesse. La cause est la suivante : Si nous diminuons la vitesse d'un objet avant une certaine force centripète et tournant autour de la terre à une certaine distance de celle-ci, cette même force sera de moins en moins appliquée sur l'objet et ainsi, ne suivra plus la trajectoire circulaire désignée initialement. Il aura ainsi tendance à s'approcher de la terre qui applique sur lui la plus grande force (selon la loi de la gravitation universelle de Newton). Ainsi, pour notre modèle, nous avons considérer qu'il n y a pas de trous d'air et que la vitesse du vent n'est pas prise en compte. Nous pouvons aussi dire que notre modèle est assez représentatif de la réalité car, à la vitesse orbitale (orbite stable), le satellite est environ à la même distance avec la surface terrestre (600 000 m) durant plusieurs périodes T. Si nous aurions pris une méthode d'approximation d'équations différentielles plus précise que la méthode d'Euler, comme par exemple les méthodes de Runge-Kutta, nous aurions obtenus un modèle naturellement plus précis.

# Discussion

Par rapport à l'importance et à l'intérêt du sujet choisi, nous croyons que celui-ci est d'une grande importance dans le domaine spatial et d'un intérêt distingué dans ce même domaine. Dans les cas où les ingénieurs du domaine spatial veulent libérer une orbite afin de remplacer un certain satellite par un autre, la trajectoire du satellite obsolète est ainsi importante afin que cette dite trajectoire soit en concordance avec les trajectoires des autres satellites (pour qu'il n y ait pas de collisions entre eux). Le temps de cette trajectoire est aussi important afin de savoir à quel moment le satellite touchera la surface terrestre. Par rapport aux modèles obtenus, nous pouvons clairement dire que cela concorde moyennement à la réalité si nous considérons l'absence de trous d'air dans l'atmosphère et que la vitesse du vent est nulle. Cette dernière contrainte rend donc notre modèle assez moyen par rapport aux autres travaux de recherches universitaires. Par rapport à ce qui reste à faire, nous croyons que le calcul de la dégradation du satellite à l'approche de la terre est une chose importante à considérer. Nous croyons aussi que prendre en compte de la vitesse du vent serait une chose à faire si nous voulons un modèle de plus en plus réaliste. Comme dernier ajout, il serait intéressant d'approximer le coefficient $C_x$ du satellite si celui-ci avait la forme de Hubble ou d'un autre satellite connu. Cette dernière étape nécessite donc des connaissances universitaires poussées.

# Bibliographie

# Annexe

Preuve de la troisième loi de Kepler :

Supposons qu'un objet de masse m tourne autour de la terre (masse M) de manière circulaire et ayant une vitesse constante v. La distance entre l'objet et le centre de la terre est r :

Alors,

$$\frac{GMm}{r^2} = \frac{mv^2}{r}$$
$$\frac{GM}{r} = v^2$$
$$v = \sqrt{\frac{GM}{r}}$$

Si nous considérons que la période est : $T = \frac{2\pi r}{v}$

Alors,

$$T = \frac{2\pi}{\sqrt{GM}}\sqrt{r^3}$$

Tableaux pour l'orbite sans frottement :

Par tableau, pour une vitesse initiale de 7561,72 m/s, nous obtenons pour x et y, en limitant le temps à 31,45 s :

| t (s) | x (m) | y (m) |
|---|---|---|
| 0 | 6956752,0000 | 0,0000 |
| 1,85 | 6956737,9025 | 13701,4885 |
| 3,7 | 6956695,6102 | 27402,9492 |
| 5,55 | 6956625,1230 | 41104,3267 |
| 7,4 | 6956526,4412 | 54805,5653 |
| 9,25 | 6956399,5652 | 68506,6095 |
| 11,1 | 6956244,4952 | 82207,4039 |
| 12,95 | 6956061,2318 | 95907,8929 |
| 14,8 | 6955849,7756 | 109608,0209 |
| 16,65 | 6955610,1273 | 123307,7324 |
| 18,5 | 6955342,2876 | 137006,9719 |
| 20,35 | 6955046,2576 | 150705,6840 |
| 22,2 | 6954722,0381 | 164403,8129 |
| 24,05 | 6954369,6303 | 178101,3033 |
| 25,9 | 6953989,0353 | 191798,0996 |
| 27,75 | 6953580,2545 | 205494,1463 |
| 29,6 | 6953143,2893 | 219189,3878 |
| 31,45 | 6952678,1411 | 232883,7687 |

Pour les mêmes valeurs de t, nous trouvons pour g en fonction de la hauteur du satellite :

| r (m) | g (m/s²) |
|---|---|
| 6956752,00 | -8,24 |
| 6956751,97 | -8,24 |
| 6956751,87 | -8,24 |
| 6956751,71 | -8,24 |
| 6956751,49 | -8,24 |
| 6956751,20 | -8,24 |
| 6956750,85 | -8,24 |
| 6956750,43 | -8,24 |
| 6956749,95 | -8,24 |
| 6956749,41 | -8,24 |
| 6956748,80 | -8,24 |
| 6956748,13 | -8,24 |
| 6956747,40 | -8,24 |
| 6956746,60 | -8,24 |
| 6956745,74 | -8,24 |
| 6956744,82 | -8,24 |
| 6956743,84 | -8,24 |
| 6956742,79 | -8,24 |

Tableaux pour une trajectoire avec frottement et une vitesse orbitale stable : Voilà, encore une fois, un tableau ayant comme temps maximum : 31,45 s, la position en x et la position en y. À noter que ceci n'est qu'un échantillon de tableau tout comme précédemment.

| t (s) | x (m) | y (m) |
|---|---|---|
| 0 | 6956752 | 0 |
| 1,85 | 6956737,903 | 13989,182 |
| 3,7 | 6956695,61 | 27978,3357 |
| 5,55 | 6956625,123 | 41967,4043 |
| 7,4 | 6956526,441 | 55956,3311 |
| 9,25 | 6956399,565 | 69945,0596 |
| 11,1 | 6956244,496 | 83933,5329 |
| 12,95 | 6956061,233 | 97921,6943 |
| 14,8 | 6955849,777 | 111909,487 |
| 16,65 | 6955610,13 | 125896,855 |
| 18,5 | 6955342,292 | 139883,741 |
| 20,35 | 6955046,265 | 153870,088 |
| 22,2 | 6954722,048 | 167855,84 |
| 24,05 | 6954369,644 | 181840,94 |
| 25,9 | 6953989,055 | 195825,331 |
| 27,75 | 6953580,28 | 209808,957 |
| 29,6 | 6953143,323 | 223791,762 |
| 31,45 | 6952678,184 | 237773,687 |

Avec les mêmes valeurs de t, nous les accélérations gravitationnelles respectives :

| t (s) | g (m/s²) |
|---|---|
| 0 | -8,23811 |
| 1,85 | -8,23811 |
| 3,7 | -8,23811 |
| 5,55 | -8,23811 |
| 7,4 | -8,23811 |
| 9,25 | -8,23811 |
| 11,1 | -8,23811 |
| 12,95 | -8,23811 |
| 14,8 | -8,23811 |
| 16,65 | -8,23811 |
| 18,5 | -8,23812 |
| 20,35 | -8,23812 |
| 22,2 | -8,23812 |
| 24,05 | -8,23812 |
| 25,9 | -8,23812 |
| 27,75 | -8,23813 |
| 29,6 | -8,23813 |
| 31,45 | -8,23813 |